\documentclass[12pt,letterpaper]{article}
\usepackage{color}
\usepackage{booktabs}
\usepackage{cite}
\definecolor{ltgrey}{rgb}{0.90,0.90,0.90}
\newcommand{\qbar}{\ensuremath{{\bar{\q}}}}
\newcommand{\DRbar}{{\ensuremath{\overline{\mathrm{DR}}}}}
\newcommand{\MSbar}{{\ensuremath{\overline{\mathrm{MS}}}}}
\newcommand{\ttt}[1]{\texttt{#1}}
\newcommand{\mrm}[1]{\mathrm{#1}}
\newcommand{\ti}[1]{\tilde{#1}}
\newcommand{\half}{\ensuremath{\textstyle \frac12}}
\newcommand{\st}{\ensuremath{^{\mathrm{st}}}}
\newcommand{\nd}{\ensuremath{^{\mathrm{nd}}}}
\newcommand{\rd}{\ensuremath{^{\mathrm{rd}}}}
\newcommand{\neut}{\ensuremath{\ti{\chi}^0}}
\newcommand{\grav}{\ensuremath{\ti{\G}}}
\newcommand{\charg}{\ensuremath{\ti{\chi}^+}}

\renewcommand{\b}{\mathrm{b}}
\renewcommand{\c}{\mathrm{c}}
\renewcommand{\d}{\mathrm{d}}
\newcommand{\e}{\mathrm{e}}

\newcommand{\g}{\mathrm{g}}
\newcommand{\hrm}{\mathrm{h}}

\newcommand{\q}{\mathrm{q}}
\newcommand{\s}{\mathrm{s}}
\renewcommand{\t}{\mathrm{t}}
\renewcommand{\u}{\mathrm{u}}
\newcommand{\A}{\mathrm{A}}

\newcommand{\G}{\mathrm{G}}
\renewcommand{\H}{\mathrm{H}}

\newcommand{\W}{{W}}
\newcommand{\Z}{{Z}}
\newcommand{\sqt}{\ensuremath{\ti{\t}}}
\newcommand{\GeV}{\mathrm{GeV}}
\newcommand{\twovec}[2]{\ensuremath{
\left(\begin{array}{c}#1\\#2\end{array}\right)}
}
\newcommand{\grbox}[1]{\colorbox{ltgrey}{#1}\color{black}}

\newlength{\captivewidth}
\newcommand{\captive}[1]{\rule{5mm}{0mm}%
\begin{minipage}{\captivewidth}%
\caption[small]{#1}\end{minipage}}

\newlength{\tablinsep}
\newlength{\halfpagewid}
\newlength{\halfpage}
\newlength{\abstwidth}
\newcommand{\snumentry}[2]{
\begin{minipage}[t]{1.2cm}\flushright\ttt{#1}\end{minipage}
\hspace{2mm}:
\begin{minipage}[t]{11cm}\noindent
#2\end{minipage}\\[1mm]
}
\newcommand{\numentry}[2]{
\begin{minipage}[t]{1.3cm}\flushright\ttt{#1}\end{minipage}
\hspace{5mm}: 
\begin{minipage}[t]{13cm}\noindent
#2\end{minipage}\\[2mm]
}

\newcommand{\arrdes}[1]{\begin{center}\framebox{\parbox{\textwidth}{#1}}%
\end{center}}

\newcommand{\mgut}{\ensuremath{M_{\mathrm{input}}}}
\newcommand{\mmess}{\ensuremath{M_{\mathrm{mess}}}}
\newcommand{\glu}{\ensuremath{{\ti{\g}}}}
\newcommand{\sqd}{\ensuremath{{\ti{\d}}}}

\newcommand{\squ}{\ensuremath{{\ti{\u}}}}

\newcommand{\sqs}{\ensuremath{{\ti{\s}}}}

\newcommand{\sqc}{\ensuremath{{\ti{\c}}}}

\newcommand{\sqb}{\ensuremath{{\ti{\b}}}}

\newcommand{\se}{\ensuremath{{\ti{\e}}}}

\newcommand{\snu}{\ensuremath{{\ti{\nu}}}}

\newcommand{\smu}{\ensuremath{{\ti{\mu}}}}

\newcommand{\stau}{\ensuremath{{\ti{\tau}}}}

\begin{document}
\vspace*{-12mm}\begin{minipage}{0.98\textwidth}\footnotesize
\flushright
hep-ph/0311123\\
LU TP 03-39, SHEP-03-24,\\
CERN-TH/2003/204, ZU-TH 15/03, \\
LMU 19/03, DCPT/03/108, IPPP/03/54\\
CTS-IISc/2003-07, DESY 03-166, MPP-2003-111
\end{minipage}\\[8mm]\hspace*{-0.025\textwidth}
\begin{minipage}{1.05\textwidth}\center{
\Large{\bf
SUSY Les Houches Accord: Interfacing SUSY Spectrum Calculators, Decay
  Packages, and Event Generators}\\[9mm]
\normalsize P.~Skands\footnotemark[1],
B.C.~Allanach$^2$,      % hep-ph & CPC OK.
H.~Baer$^3$,            % hep-ph & CPC OK.
C.~Bal\'azs$^{3,4}$,        % hep-ph & CPC OK.  
G.~B\'elanger$^2$,
F.~Boudjema$^2$, 
A.~Djouadi$^{5,6}$, 
R.~Godbole$^7$, 
J.~Guasch$^8$,
S.~Heinemeyer$^{6,9}$,      % hep-ph & CPC OK.
W.~Kilian$^{10}$,          % hep-ph & CPC OK.
J-L.~Kneur$^{5}$,       % hep-ph & CPC OK.
S.~Kraml$^6$,
F.~Moortgat$^{11}$,     % hep-ph & CPC OK.
S.~Moretti$^{12}$,      % hep-ph & CPC OK. 
M.~M\"uhlleitner$^{8}$, % hep-ph & CPC OK. 
W.~Porod$^{13}$,        % hep-ph & CPC OK. 
A.~Pukhov$^{14}$,       % CPC OK.
P.~Richardson$^{6,15}$,
S.~Schumann$^{16}$,     % CPC OK. 
P.~Slavich$^{17}$,      % CPC OK.
M.~Spira$^{8}$, 
G.~Weiglein$^{15}$%
\\[8mm]
\begin{minipage}{14.5cm}
\small\it
\begin{tabular}{rp{13.3cm}}
$^{\ 1}$&%Theoretical Physics, 
Theoretical Physics, Lund University, S\"olvegatan 14A, 22362 Lund, Sweden.\\
$^{\ 2}$&LAPTH, 9 Chemin de Bellevue, BP 110 74941 Annecy-le-Vieux, Cedex,
France.\\ 
$^{\ 3}$&%Department of Physics
Dept.~of Physics, Florida State University, Tallahassee, FL 32306, U.S.A.\\
%Dept.~of Physics, 
%University of Hawaii, 2505 Correa Rd., Honolulu, Hawaii 96822.\\
$^{\ 4}$&
HEP Division,\\
& Argonne National Laboratory, 9700 Cass Ave., Argonne, IL 60439,
U.S.A.\\
$^{\ 5}$&%LPMT, UMR 5825 CNRS, 
LPMT, Universit\'e Montpellier II, 34095 Montpellier, Cedex 5, France.\\
$^{\ 6}$&Theory Division, 
CERN, CH--1211 Geneva 23, Switzerland.\\
$^{\ 7}$&%Centre for Theoretical Studies, 
CTS, Indian Institute of Science, Bangalore, 560012, India.\\
$^{\ 8}$&Paul Scherrer Institut, CH--5232 Villigen
PSI, Switzerland.\\
$^{\ 9}$&LMU M\"unchen, Theresienstr.~37, D--80333 M\"unchen, Germany.\\
$^{10}$&Theory Group, DESY, D--22603 Hamburg, Germany.\\
$^{11}$&University of Antwerpen, B--2610 Antwerpen, Belgium.\\
$^{12}$&%School of Physics \& Astronomy, 
School of Physics and Astronomy,\\ 
&University of Southampton, Highfield, Southampton SO17 1BJ, UK.
\\ 
$^{13}$&Institute for Theoretical Physics,\\ 
&University of Z\"urich, Winterthurerstr.~190, CH--8057 Z\"urich,
Switzerland.\\ 
$^{14}$&Moscow State University, Russia.\\
$^{15}$&IPPP, University of Durham, Durham DH1 3LE, UK.\\
$^{16}$&Institut f\"ur Theoretische Physik, TU Dresden, D--01062 Dresden, Germany.\\
$^{17}$&Max Planck Institut f\"ur Physik, F\"ohringer Ring 6, D--80805 M\"unchen, Germany\\
\end{tabular}
\end{minipage}\\[5mm]}
{November 2009}\\[4mm]
\begin{abstract}
An accord specifying a unique set of conventions for supersymmetric
extensions of the Standard Model together with generic file structures for 
1) supersymmetric model specifications and input parameters, 
2) electroweak scale supersymmetric mass and coupling spectra, and 
3) decay tables is presented, to provide a universal interface between
spectrum calculation programs, decay packages, and high energy physics
event generators. 
\end{abstract}
%\raisebox{-5mm}{\footnotesize$^1$peter.skands@thep.lu.se}
\end{minipage}
\footnotetext[1]{skands@fnal.gov. See 
  \texttt{home.fnal.gov/$\sim$skands/slha/} for updates and examples.}
\clearpage
\pagestyle{plain}\noindent
\rule{\textwidth}{0.7mm}\vspace*{-0.7cm}
\tableofcontents\vspace*{1mm}
\noindent\rule{\textwidth}{0.7mm}
\section{Introduction}
An increasing number of advanced programs for the calculation of the
supersymmetric (SUSY) mass and coupling spectrum are appearing
\cite{Allanach:2001kg,Baer:1993ae,Djouadi:2002ze,Heinemeyer:1998yj,Porod:2003um}  
in step with the more and more refined approaches which are taken in the
literature. Furthermore, these programs are often interfaced to
specialized decay packages
\cite{Beenakker:1996ed,Djouadi:1998yw,Heinemeyer:1998yj,Muhlleitner:2003vg,Porod:2003um}, 
relic density calculations 
\cite{Belanger:2001fz,Gondolo:2002tz}, and (parton--level) event generators
\cite{Baer:1999sp,Corcella:2000bw,Gleisberg:2003xi,Pukhov:1999gg,Tanaka:1997qn,Sjostrand:2006za,Kilian:2007gr}, 
in themselves fields with a proliferation of philosophies and,
consequentially, programs.

At 
present, a small number of specialized interfaces exist between various
codes. Such tailor-made interfaces are not easily generalized
and are time-consuming to construct and test
for each specific implementation. A universal interface would 
 clearly be an advantage here.  

However, since the 
codes involved are not all written in the same
programming language, 
the question naturally arises how to make such an interface work across
 languages. At present, an inter-language 
linking solution 
 does not seem to be feasible without introducing at least some dependence on
 platform (e.g.\ 
 UNIX variant) and/or compiler. For details on these aspects, see e.g.\
 \cite{cnl94}.  

At this point, we deem such an interface too fragile to be set loose
among the particle physics community. 
Instead, we advocate a less elegant but more robust
solution, exchanging information between FORTRAN and C(++) codes
via three ASCII files, one for model input, one for model input plus 
spectrum output, and one for model input plus spectrum output plus decay
information. The detailed structure of these files is described in the
sections below. 
Briefly stated, the purpose of this Accord is thus the
following: 
\begin{enumerate}
\item To present 
a set of generic definitions for an input/output file structure which 
provides a universal framework for interfacing SUSY spectrum calculation
programs.
\item To present a generic file structure for the transfer of decay
  information between decay calculation packages and event generators.
\end{enumerate}
\begin{figure}[h]
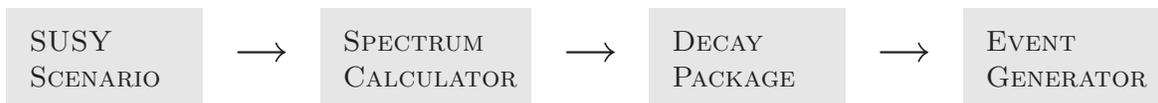

{\small
\begin{center}
\begin{tabular}{lclclcl}
\grbox{\sc\begin{minipage}{2cm}\sc SUSY\\Scenario\end{minipage}} & 
{\LARGE ${\mathbf\rightarrow}$}& 
\grbox{\sc\begin{minipage}{2.2cm}
Spectrum\\
Calculator\end{minipage}} &
{\LARGE${\mathbf\rightarrow}$}& 
\grbox{\sc\begin{minipage}{2cm}
Decay\\Package\end{minipage}}&
{\LARGE${\mathbf\rightarrow}$}& 
\grbox{\sc\begin{minipage}{2.1cm}Event\\Generator
\end{minipage}}\\
\end{tabular}
\caption{Stages of the interface Accord. By SUSY scenario we not only 
intend specific SUSY--breaking mechanisms, such as supergravity (SUGRA), gauge
mediated SUSY breaking (GMSB), etc., but also 
more general setups within the Minimal Supersymmetric Standard Model
(MSSM). \label{fig:codes}} 
\end{center}}
\end{figure}
Note that different codes may have different implementations of 
how SUSY Les Houches Accord (SLHA) input/output is \emph{technically} 
achieved. The details of how to `switch on' SLHA input/output with a
particular program should be
described in the manual of that program and are not covered here.
%This also has the advantage over a run-time interface that the user may
%perform his or her own consistency checks at intermediate stages in the
%calculation, e.g.\ checking the spectrum file to see whether a mass spectrum
%looks as expected. 

\subsection{Using the Accord}
To interface two or more calculations, the general procedure would be that
the user prepares model input parameters together with a set of Standard
Model parameters (to be used as low--scale boundary conditions for the
spectrum calculation) in an ASCII file, complying with the standard defined
in section \ref{sec:input} below. At present, only models with the particle
spectrum of the MSSM, and with CP and R--parity conserved are included in
this standard.

The user then runs a spectrum calculation program
%(henceforth denoted RGE package) 
with these inputs to obtain the SUSY mass and coupling spectrum at the
electroweak (EW) scale. The resulting spectrum is stored, together with a
copy of the model input parameters (so that subsequent calculations may be
performed consistently), in a new file. Standards for the spectrum file are
defined in section \ref{sec:spectrum}.

The user may now run some particular decay package to generate a list of
decay modes and widths for selected particles, which the decay package saves
to a third file, complying with the definitions in section
\ref{sec:decay}. Again, a copy of the model input parameters as well as the
complete spectrum information is included together with the decay information
in this file.

Lastly, the user may instruct a (parton--level) event generator to read in
all this information and start generating events. Of course, any of these
intermediate steps may be skipped whenever the user does not wish to switch
between programs across them, e.g.\ no decay information is required to be
present in the file read by an event generator if the user wishes the event
generator to calculate all decay widths itself.

If a general purpose event generator is used, the events will include parton
showering and hadronization, whereas if a parton--level generator is used,
the events may finally be passed to parton showering and hadronization
programs using the already defined Les Houches Accord \#1
\cite{Boos:2001cv,Alwall:2006yp}.

\section{Conventions \label{sec:conventions}}
One aspect of supersymmetric calculations that has often given rise to
confusion and consequent inconsistencies in the past is the multitude of ways
in which the parameters can be, and are being, defined. Hoping to minimize
both the extent and impact of such confusion, we have chosen to adopt one
specific set of self-consistent conventions for the parameters appearing in
this Accord. These conventions are described in the following subsections. As
yet, we only consider R--parity and CP conserving scenarios, with the particle
spectrum of the MSSM.

Compared to the widely used Gunion and Haber conventions and notation
\cite{Gunion:1986yn}, our prescriptions show a few differences.  These will
be remarked upon in all places where they occur below, with parameters in the
notations and conventions of \cite{Gunion:1986yn} denoted with an explicit
superscript: $^\mrm{GH}$.

\subsection{Standard Model Parameters \label{sec:smconv}}
In general, the SUSY spectrum calculations impose low--scale boundary
conditions on the renormalization group equation (RGE) flows to ensure that
the theory gives correct predictions for low--energy observables. Thus,
experimental measurements of masses and coupling constants at the electroweak
scale enter as inputs to the spectrum calculators.

In this Accord, we choose a specific set of low--scale input parameters
(block \texttt{SMINPUTS} below), letting the electroweak sector be fixed by
\begin{enumerate} 
\item $\alpha_\mrm{em}(m_\Z)^{\MSbar}$: the electromagnetic
  coupling at the $\Z$ pole in the \MSbar\ scheme with 5 active flavours 
(see e.g.~\cite{Hagiwara:2002fs}\footnote{Note that \cite{Hagiwara:2002fs}
  uses the notation $\widehat{\alpha}(M_\Z)$ for this parameter.}).
This coupling is connected to the classical 
fine structure constant, $\alpha=
1/137.0359895(61)$ \cite{Hagiwara:2002fs} by a relation of the form:
\begin{equation}
\alpha_\mrm{em}(m_\Z)^{\MSbar} = \frac{\alpha}{1 -
   \Delta\alpha(m_\Z)^\MSbar},
\end{equation}
where $\Delta\alpha(m_\Z)^\MSbar$ contains the quantum corrections involved
in going from the classical limit to the \MSbar\ value at the scale
$m_\Z$. 
\item $G_F$: the Fermi constant determined from muon decay.
\item $m_\Z$: the $\Z$ boson pole mass. 
\end{enumerate} 
All other electroweak parameters, such as $m_W$ and $\sin^2\theta_W$, should
be derived from these inputs if needed. 

The strong interaction strength is fixed by $\alpha_s(m_\Z)^{\MSbar}$ (the
five--flavour strong coupling at the scale $m_\Z$ in the \MSbar\ scheme), and
the third 
generation Yukawa couplings are obtained from the top and tau pole masses,
and from $m_b(m_b)^{\MSbar}$, see \cite{Hagiwara:2002fs}. 
The reason we take the running $b$ mass in the
\MSbar\ scheme rather than a pole mass definition is that the latter suffers
from infra-red sensitivity problems, hence the former is the quantity which
can be most accurately related to experimental measurements. If required,
relations between running and pole quark masses may be found in
\cite{Melnikov:2000qh,Baer:2002ek}. 
 
It is also important to note that all presently available experimental
determinations of e.g.~$\alpha_s$ and the running $b$ mass are based on
assuming the Standard Model as the underlying theory, for natural
reasons. When extending the field content of the SM to that of the MSSM, the
\emph{same} measured results would be obtained for \emph{different} values of
these quantities, due to the different underlying field content present in
the MSSM.  However, since these values are not known before the spectrum has
been determined, all parameters contained in block \texttt{SMINPUTS} should
be the `ordinary' ones obtained from SM fits, i.e.~with no SUSY corrections
included.  The spectrum calculators themselves are then assumed to convert
these parameters into ones appropriate to an MSSM framework.

Finally, while we assume $\MSbar$ running quantities with the SM as the
underlying theory as input, all running parameters in the \emph{output} of
the spectrum calculations are defined in the modified dimensional reduction
($\DRbar$) scheme \cite{Siegel:1979wq,Capper:1980ns,Jack:1994rk}, with
different spectrum calculators possibly using different prescriptions for the
underlying effective field content. More on this in section
\ref{sec:regularization}.

\subsection{Supersymmetric Parameters \label{sec:susypar}}
The chiral superfields of the MSSM have the following $SU(3)_C\otimes
SU(2)_L\otimes U(1)_Y$ quantum numbers
\begin{eqnarray}
L:&(1,2,-\half),\quad {\bar E}:&(1,1,1),\qquad\, \textstyle
Q:\,(3,2,\frac16),\quad
{\bar U}:\,(\bar{3},1,-\frac{2}{3}),\nonumber\\ {\bar D}:&(\bar{3},1,\frac13),\quad
H_1:&(1,2,-\half),\quad  H_2:\,(1,2,\half)~.
\label{fields}
\end{eqnarray}
Then, the superpotential (omitting RPV terms) is written as
\begin{eqnarray}
W&=& \epsilon_{ab} \left[ 
  (Y_E)_{ij} H_1^a    L_i^b    {\bar E}_j 
+ (Y_D)_{ij} H_1^a    Q_i^{b}  {\bar D}_{j} 
+ (Y_U)_{ij} H_2^b    Q_i^{a}  {\bar U}_{j}  
- \mu H_1^a H_2^b \right]~.
\label{eq:superpot}
\end{eqnarray}

Throughout this section, we denote $SU(2)_L$ fundamental representation
indices by $a,b=1,2$ and generation indices by $i,j=1,2,3$. Colour indices
are everywhere suppressed, since only trivial contractions are involved.
$\epsilon_{ab}$ is the totally antisymmetric tensor, with
$\epsilon_{12}=\epsilon^{12}=1$. Lastly, we will use ${t,b,\tau}$ to
denote the $i=j=3$ entries of mass or coupling matrices (top, bottom and tau).

The Higgs vacuum expectation values (VEVs) are $\langle H_i^0 \rangle =
v_i/\sqrt{2}$, and $\tan\beta=v_2/v_1$. We also use the notation
$v=\sqrt{v_1^2+v_2^2}$. Different choices of 
renormalization scheme and scale are possible for defining $\tan\beta$. For
the input to the spectrum calculators, we adopt by default the commonly
encountered definition 
\begin{equation}
\tan\beta(m_\Z)^{\DRbar},
\end{equation}
i.e.~the $\tan\beta$ appearing in block \ttt{MINPAR} below is
defined as a \DRbar\ running parameter given at the scale $m_\Z$. The optional
extended input block \ttt{EXTPAR} allows the  
possibility of using an input definition at a different scale,
$\tan\beta(\mgut\ne m_\Z)^{\DRbar}$. Lastly, the spectrum calculator may be
instructed to write out one or several values of $\tan\beta(Q)^{\DRbar}$ at
various scales $Q_i$, see section \ref{sec:regularization} and block
\ttt{HMIX} below. 

The MSSM \DRbar\ gauge couplings (block \ttt{GAUGE} below) are: $g'$
(hypercharge gauge coupling in Standard Model normalization), $g$ ($SU(2)_L$
gauge coupling) and $g_3$ (QCD gauge coupling).
 
Our Yukawa matrices, $Y_E,~Y_D$, and $Y_U$, correspond exactly to
$(f)^{\mrm{GH}},~(f_1)^{\mrm{GH}}$, and $(f_2)^{\mrm{GH}}$, respectively, in
the notation of \cite{Gunion:1986yn}. For hypercharge, \cite{Gunion:1986yn}
uses $(y)^{\mrm{GH}} \equiv 2Y$, and for the $SU(2)_L$ singlet leptonic
superfield the notation $(\hat{R})^{\mrm{GH}}\equiv\bar{E}$. Finally, for the
Higgs vacuum expectation values, we choose the convention in which $\langle
H_i^0 \rangle = v_i/\sqrt{2} \equiv (v_i)^{\mrm{GH}}$, so that
$v^2 = (v_1^2+v_2^2) = (246~\GeV)^2 $, corresponding to $m_\Z^2=\frac14
(g'^2+g^2)(v_1^2+v_2^2)$, whereas \cite{Gunion:1986yn} has
$(v_1^2+v_2^2)^\mrm{GH} = (174~\GeV)^2$, with $m_\Z^2=\frac12
(g'^2+g^2)(v_1^2+v_2^2)^{\mrm{GH}}$. Otherwise, conventions for the
superpotential are identical between this article and
\cite{Gunion:1986yn}\footnote{The sign of $\mu$ in the original Gunion and
Haber article actually disagrees with ours, but that sign was subsequently
changed in the erratum to that article, which we here include when giving
reference to \cite{Gunion:1986yn}.}.

\subsection{SUSY Breaking Parameters \label{sec:susybreak}}

We now tabulate the notation of the soft SUSY breaking parameters.
The trilinear scalar interaction potential is
\begin{equation}
V_3 = \epsilon_{ab} \sum_{ij}
\left[
(T_E)_{ij} H_1^a \tilde{L}_{i_L}^{b} \tilde{e}_{j_R}^* +
(T_D)_{ij} H_1^a               \tilde{Q}_{i_L}^{b}  \tilde{d}_{j_R}^* +
(T_U)_{ij}  H_2^b \tilde{Q}_{i_L}^{a} \tilde{u}_{j_R}^*
\right]
+ \mrm{h.c.}~,
\label{eq:trilinear} 
\end{equation}
where fields with a tilde are the scalar components of the superfield
with the identical capital letter  (note however that we define, e.g.,
$\tilde{u}_R^*$ as the scalar component of $\bar{U}$). 
In the literature the T matrices are often decomposed as 
\begin{equation}
\frac{T_{ij}}{Y_{ij}} = A_{ij}~~~~~; (\mathrm{no~sum~over~}i,j)~,
\end{equation}
where $Y$ are the Yukawa matrices and $A$ the soft supersymmetry breaking
trilinear couplings. See also blocks \ttt{YE}, \ttt{YD},
\ttt{YU}, \ttt{AE},  \ttt{AD}, and \ttt{AU} below.
 
The scalar bilinear SUSY breaking terms are contained in the potential
\begin{eqnarray}
V_2 &=& m_{H_1}^2 {{H^*_1}_a} {H_1^a} + m_{H_2}^2 {{H^*_2}_a} {H_2^a} +
{\tilde{Q}^*}_{i_La} (m_{\tilde Q}^2)_{ij} \tilde{Q}_{j_L}^{a} +
{\tilde{L}^*}_{i_La} (m_{\tilde L}^2)_{ij} \tilde{L}_{j_L}^{a}  
+ \nonumber \\ &&
\tilde{u}_{i_R} (m_{\tilde u}^2)_{ij} {\tilde{u}^*}_{j_R} +
\tilde{d}_{i_R} (m_{\tilde d}^2)_{ij} {\tilde{d}^*}_{j_R} +
\tilde{e}_{i_R} (m_{\tilde e}^2)_{ij} {\tilde{e}^*}_{j_R} -
(m_3^2 \epsilon_{ab} H_1^a H_2^b + \mrm{h.c.})~. \label{eq:v2}
\end{eqnarray}
Rather than using $m_3^2$ itself, below we use the more convenient 
parameter $m_A^2$, defined by:
\begin{equation}
m_A^2 = \frac{m_3^2}{\sin\beta\cos\beta}, \label{eq:mA}
\end{equation}
which is identical to the pseudoscalar Higgs mass at tree level in our
conventions. We emphasize, however, that this parameter is solely defined by
eq.~(\ref{eq:mA}) and should not be confused with the pole mass of the
$\A^0$, $m_{\A^0}$, which is an alternative possible input
parameter and which has its own separate entry in this accord. 

Writing the bino as ${\tilde b}$, the
unbroken $SU(2)_L$ gauginos as ${\tilde w}^{A=1,2,3}$, and the gluinos as
${\tilde g}^{X=1\ldots8}$, the gaugino mass terms (appearing in blocks
\ttt{EXTPAR} and \ttt{GAUGE} below) are contained in the
Lagrangian 
\begin{equation}
{\mathcal L}_G = \frac{1}{2} \left( M_1 {\tilde b}{\tilde b} + M_2 {\tilde
    w}^A{\tilde w}^A 
+ M_3 {\tilde g}^X {\tilde g}^X \right) + \mrm{h.c.}~\label{eq:LG}.
\end{equation}

For the soft trilinear breaking terms above, we use the same sign convention as
\cite{Gunion:1986yn}, but with a different normalization and unit $(m_6
A_i)^{\mrm{GH}}\equiv A_i$. For the bilinear breaking terms, we differ only in 
notation; $(m_{i})^{\mrm{GH}}\equiv m_{H_i}$, $(m_{12})^{\mrm{GH}}\equiv m_3$,
$(M')^{\mrm{GH}} \equiv M_1$, $(M)^{\mrm{GH}} \equiv M_2$,
$(\lambda')^{\mrm{GH}}\equiv \tilde{b}$, and $(\lambda^A)^{\mrm{GH}}\equiv
\tilde{w}^A$. Below, it will also be useful to note that we use $\tilde{h}_1
\equiv (\psi^0_{H_1})^\mrm{GH}$ and $\tilde{h}_2 \equiv
(\psi^0_{H_2})^\mrm{GH}$ for the higgsinos. 

\subsection{Mixing Matrices \label{sec:mixing}}
In the following, we describe in detail our conventions for neutralino,
chargino, sfermion, and Higgs mixing. For purposes of cross-checking, 
we include in Appendix \ref{app:mixing} expressions for
the tree--level mass matrices of neutralinos, charginos, and third generation
sfermions, as they appear in the set of conventions adopted here.  

More importantly, essentially all SUSY spectrum calculators 
on the market today work with
mass matrices which include higher--order corrections. 
Consequentially, a formal dependence on the renormalization scheme and scale,
and on the external momenta appearing in the corrections, 
enters the definition of the corresponding mixing matrices. Since, at the
moment, no consensus exists on the most convenient definition to use here, the
parameters appearing in blocks \ttt{NMIX}, \ttt{UMIX}, \ttt{VMIX}, 
\ttt{STOPMIX}, \ttt{SBOTMIX}, \ttt{STAUMIX}, and \ttt{ALPHA} below should be
thought of as `best choice' solutions, at the discretion of each spectrum
calculator. For example, one program may output on--shell
parameters (with the external momenta e.g.~corresponding to specific particle
masses) in these blocks while another may be using $\DRbar$ definitions at
certain `characteristic' scales. For details on specific prescriptions, 
the manual of the particular spectrum calculator should be consulted. 

Nonetheless, for obtaining loop--improved tree--level results, 
these parameters can normally be used as is. They can 
also be used for consistent cross section and decay width calculations
at higher orders, but then the renormalization prescription
employed by the spectrum calculator must match or be consistently matched to
that of the intended higher order calculation.

Finally, different spectrum calculators may disagree on the overall sign of
one or more rows in a mixing matrix, owing to different 
diagonalization algorithms. Such differences 
correspond to a flip of the sign of the eigenvectors in question and do
not lead to inconsistencies. Only the relative sign between entries on the
same row is physically significant, for processes with interfering amplitudes.

\subsubsection{Neutralino Mixing \label{conv:nmix}} 
The Lagrangian contains the (symmetric) neutralino mass matrix as 
\begin{equation}
\mathcal{L}^{\mrm{mass}}_{\neut} =
-\frac12{\tilde\psi^0}{}^T{\cal M}_{\tilde\psi^0}\tilde\psi^0 +
\mathrm{h.c.}~, 
\end{equation}
in the basis of 2--component spinors $\tilde\psi^0 =$
$(-i\tilde b,$ $-i\tilde w^3,$ 
$\tilde h_1,$ $\tilde h_2)^T$. We define the unitary 4 by 4  
neutralino mixing matrix $N$ (block \ttt{NMIX} below), such that:
\begin{equation}
-\frac12{\tilde\psi^0}{}^T{\cal M}_{\tilde\psi^0}\tilde\psi^0
= -\frac12\underbrace{{\tilde\psi^0}{}^TN^T}_{{\neut}{}^T} \underbrace{N^*{\cal
    M}_{\tilde\psi^0}N^\dagger}_{\mathrm{diag}(m_{\neut})}
\underbrace{N\tilde\psi^0}_{\neut}~,  \label{eq:neutmass}
\end{equation}
where the (2--component) neutralinos $\neut_i$ are defined such that their
absolute masses 
increase with increasing $i$. Generically, the resulting mixing matrix $N$
may yield complex entries in the mass matrix,
$\mathrm{diag}(m_{\neut})_i=m_{\neut_i}e^{i\varphi_i}$. If so, 
we absorb the phase into the definition of the corresponding eigenvector,
$\neut_i \to \neut_i e^{i\varphi_i/2}$, making
the mass matrix strictly real: 
\begin{equation}
\mathrm{diag}(m_{\neut})\equiv\left[N^*
  {\cal M}_{\tilde\psi^0} N^\dagger\right]_{ij}=m_{\neut_i}\delta_{ij}. 
\end{equation}
Note, however, that a
special case occurs when CP violation is absent and one or more of the
$m_{\neut_i}$ turn out to be negative. In this case, we allow for maintaining 
a strictly real mixing matrix $N$, instead writing the \emph{signed} mass
eigenvalues in the output. 
Thus, a negative $m_{\neut_i}$ in the output implies that the
physical field is obtained by the rotation $\neut_i \to \neut_i e^{i\pi_i/2}$.

Our conventions on this point are slightly different from those used in
\cite{Gunion:1986yn} where the same Lagrangian appears but where the
diagonalizing matrix $(N)^\mrm{GH}$ is
chosen such that the elements of the diagonal mass matrix,
$(N_D)^\mrm{GH}$ (the mass matrix, in the notation of \cite{Gunion:1986yn}), 
are \emph{always} real and non-negative, i.e.~$(N_D)^\mrm{GH}\equiv
|\mathrm{diag}(m_{\neut})|$, at the price of $(N)^\mrm{GH}$ generally 
being complex--valued also in the absence of CP violation.

\subsubsection{Chargino Mixing \label{conv:chmix}} 
We make the identification ${\tilde w}^\pm = ({\tilde w^1} \mp i{\tilde w^2}
) / \sqrt{2}$ for the charged winos and ${\tilde h_1^-}, {\tilde h_2^+}$ for
the charged higgsinos.  The Lagrangian contains the chargino mass matrix as
\begin{equation}
\mathcal{L}^{\mrm{mass}}_{\charg} = -{\tilde\psi^-}{}^T{\cal M}_{\tilde\psi^+}\tilde\psi^+ +
\mrm{h.c.}~, 
\end{equation}
in the basis of 2--component spinors $\tilde\psi^+ = (-i\tilde
w^+,\ \tilde h_2^+)^T,\ \tilde\psi^-= (-i\tilde w^-,\ \tilde h_1^-)^T$.  We
define the unitary 2 by 2 chargino mixing matrices, $U$ and
$V$ (blocks \ttt{UMIX} and \ttt{VMIX} below), such that:
\begin{equation}
-{\tilde\psi^-}{}^T{\cal M}_{\tilde\psi^+}\tilde\psi^+
= -
  \underbrace{{\tilde\psi^-}{}^TU^T}_{{\ti{\chi}^-}{}^T} 
  \underbrace{U^*{\cal M}_{\tilde\psi^+}V^\dagger}_{\mathrm{diag}(m_{\charg})}
  \underbrace{V\tilde\psi^+}_{\charg}~,  \label{eq:chargmass}
\end{equation} 
where the (2--component) charginos $\charg_i$ are defined such that their
absolute masses increase with increasing $i$ and such that the mass matrix,
$m_{\charg_i}$, is strictly real: 
\begin{equation}
\mathrm{diag}(m_{\charg})\equiv\left[U {\cal M}_{\tilde\psi^+}
  V^T\right]_{ij} = m_{\charg_i}\delta_{ij}~. 
\end{equation}
Again, in the absence of CP violation $U$ and $V$ can be chosen strictly real. 
This choice, as compared to that adopted in \cite{Gunion:1986yn}, shows
similar differences as for neutralino mixing. 

\subsubsection{Sfermion Mixing \label{conv:sfmix}}
At present, we restrict our attention to left--right mixing in the third
generation sfermion sector only.  The convention we use is, for the
interaction eigenstates, that $\ti{f}_L$ and $\ti{f}_R$ refer to the $SU(2)_L$
doublet and singlet superpartners of the fermion $f\in\{t,b,\tau\}$,
respectively, and, for the mass eigenstates, that $\ti{f}_1$ and $\ti{f}_2$
refer to the lighter and heavier mass eigenstates, respectively. With this
choice of basis, the spectrum output (blocks \ttt{STOPMIX}, \ttt{SBOTMIX},
and \ttt{STAUMIX} below) should contain the elements of the following matrix:
\begin{equation}
\twovec{{\ti{f}_1}}{{\ti{f}_2}} = \left[
\begin{array}{cc}
F_{11} & F_{12} \\
F_{21} & F_{22}
\end{array}\right]\twovec{{\ti{f}_L}}{{\ti{f}_R}}~,
\end{equation}
whose determinant should be $\pm 1$. We here deliberately avoid notation
involving mixing angles, to prevent misunderstandings which could arise due
to the different conventions for these angles used in the literature.  The
mixing matrix elements themselves are unambiguous, apart from the overall
signs of rows in the matrices, see above.  Note that in \cite{Gunion:1986yn},
the mass eigenstates are \emph{not} necessarily ordered in mass.

\subsubsection{Higgs Mixing \label{conv:hmix}}
The conventions for $\mu$, $v_1$, $v_2$, $v$, $\tan\beta$, and $m_A^2$ were
defined above in sections   
\ref{sec:susypar} and \ref{sec:susybreak}. The angle 
$\alpha$ (block \ttt{ALPHA}) we define by the rotation matrix:
\begin{equation}
\twovec{{{H}^0}}{{{h}^0}} = \left[
\begin{array}{cc}
\cos\alpha & \sin\alpha \\
-\sin\alpha & \cos\alpha
\end{array}\right]\twovec{{{H}^0_1}}{{{H}^0_2}} ~,
\end{equation}
where ${H^0_1}$ and ${H^0_2}$ are the CP--even neutral Higgs scalar
interaction eigenstates, and $h^0$ and $H^0$ the corresponding mass
eigenstates (including any higher order corrections present in the spectrum
calculation), with $m_{h^0}<m_{H^0}$ by definition. This convention is
identical to that of \cite{Gunion:1986yn}.

\subsection{Running Couplings \label{sec:regularization}}
In contrast to the effective definitions adopted above for the mixing
matrices, we choose to define the parameters which appear in the output blocks
\ttt{HMIX}, \ttt{GAUGE}, \ttt{MSOFT}, 
\ttt{AU}, \ttt{AD}, \ttt{AE}, \ttt{YU}, \ttt{YD}, and
\ttt{YE}, as \DRbar~running parameters, computed at one or more
user--specifiable scales $Q_i$.

That the \DRbar\ scheme is adopted for the output of running parameters is
simply due to the fact that this scheme substantially simplifies many SUSY
calculations (and hence all spectrum calculators use it).
However, it does have drawbacks which for some applications are serious.  For
example, the \DRbar\ scheme violates mass factorization as used in QCD
calculations~\cite{Beenakker:1988bq}. For consistent calculation beyond
tree--level of processes relying on this factorization, e.g.~cross sections
at hadron colliders, the \MSbar\ scheme is the only reasonable choice. At the
present level of calculational precision, this is fortunately not an
obstacle, since at one loop, a set of parameters calculated in either of the
two schemes can be consistently translated into the other
\cite{Martin:1993yx}. Explicit prescriptions for how to do this are included
in appendix~\ref{app:translations}.

Note, however, 
that different spectrum calculators use different choices for the
underlying particle content of the effective theory. The programs
\textsc{Softsusy} (v.~1.8), \textsc{SPheno} (v.~2.1), and \textsc{Suspect}
(v.~2.2) use the full MSSM spectrum at all scales, whereas in \textsc{Isajet}
(v.~7.69) an alternative prescription is followed, with different particles
integrated out of the effective theory at different scales. Whatever the
case, these couplings should \emph{not} be used `as is' in calculations
performed in another renormalization scheme or where a different effective
field content is assumed. Thus,
unfortunately, 
ensuring consistency of the field content assumed in the
effective theory must still be done on a per program basis, though
information on the prescription used by a particular spectrum calculator may
conveniently be given in block \ttt{SPINFO}, when running parameters are
provided.

Technically, we treat running parameters in the output in the following
manner: since programs outside the spectrum calculation will not normally be
able to run parameters with the full spectrum included, or at least less
precisely than the spectrum calculators themselves, an option is included in
block \ttt{MODSEL} below to
instruct the spectrum calculator to write out values for each running parameter
at a user--defined number of logarithmically spaced scales, i.e.\ to give
output on running parameters at a grid of scales, $Q_i$, where the lowest
point in the grid will normally be $Q_{\mrm{min}}=m_\Z$ and the highest point
is 
user--specifiable. A complementary possibility is to let the spectrum
calculator give output for the running couplings at one or more scales equal
to specific sparticle masses in the spectrum. This option is also invoked
using block \ttt{MODSEL}.

Warning: please note that these options are merely intended to \emph{allow}
information on running parameters to be passed, if desired. Many of the 
codes involved will at present not actually make use of this information,
even if provided with it.

\section{Definitions of the Interfaces}
In this section, the SUSY Les Houches Accord input and
output files are described. We here concentrate on the technical structure
only. The reader should consult section \ref{sec:conventions} for
parameter definitions and convention choices.

The following general structure for the SLHA files is proposed:
\begin{itemize}
\item All quantities with dimensions of energy (mass) are implicitly
  understood to be in GeV (GeV$/c^2$).
\item Particles are identified by their PDG particle codes. See appendix 
  \ref{app:pdg} for lists of these, relevant to the MSSM. 
For a complete listing, see \cite[chp.~33]{Hagiwara:2002fs}.
\item The first character of every line is reserved for control and
  comment statements. Data lines should have the first character empty.
\item In general, formatted output should be used for write-out, to avoid
  ``messy-looking'' files, while a free format should be used on read-in, to
  avoid misalignment etc.~leading to program crashes. 
\item Read-in should be performed in a case-insensitive way, again to
  increase stability. 
\item The general format for all real numbers is the FORTRAN format
  E16.8\footnote{E16.8: 
a 16-character wide real number in scientific notation, whereof
  8 digits are decimals, e.g.\ ``\ttt{-0.12345678E+000}''.}. 
  This large number of digits is used to avoid any possible numerical
  precision issue, and since it is no more difficult for e.g.\ the spectrum
  calculator to write out such a number than a shorter version. For typed
  input, it merely means that at least 16 spaces are reserved for the number,
  but e.g.\ the number \ttt{123.456} may be typed in ``as is''. See also the
  example files in appendix \ref{app:examples}.
\item A ``\ttt{\#}'' 
mark anywhere means that the rest of the line is intended as a comment
to be ignored by the reading program. 
\item All input and output is divided into sections in the form of named
  ``blocks''. A ``\ttt{BLOCK xxxx}'' (with the ``\ttt{B}'' being the first
character on the line) marks the beginning of entries belonging to 
the block named ``\ttt{xxxx}''. E.g.\
``\ttt{BLOCK MASS}'' marks that all following lines until the next
``\ttt{BLOCK}'' (or ``\ttt{DECAY}'') statement contain mass values, to be read 
in a specific format, intrinsic to the \ttt{MASS} block. The order of blocks
is arbitrary, except that input blocks should always come before output
blocks. 
\item
Reading programs should skip over blocks that are not recognized, issuing a
warning rather than crashing. Thereby, stability is increased and 
private blocks can be constructed, for instance \ttt{BLOCK MYCODE} could
contain some parameters that only the program \textsc{MyCode} (or a special
hack of 
it) needs, but which are not recognized universally.
\item A line with a blank first character is a data statement, to be
  interpreted according to what data the current
  block contains. Comments and/or descriptions added after the data values, 
e.g.\ ``\ttt{\ ... \# comment}'', should always be
added, to increase readability of the file for human readers. 
\item Use of the `tab' character is dangerous and should be avoided.
\end{itemize}
Finally, program authors are advised to check that any parameter relations 
they assume in their codes (implicit or explicit) are either obeyed by the
parameters in the files or disabled. As a specific example, take a
code that normally would use e.g.\ the tree--level expression for the
stop mixing matrix to 
compute $A_t$, given the stop mixing angle (together with a given set of other
input parameters). This relation should \emph{not} be used when reading in 
an SLHA spectrum; there may be (higher--order) contributions included in the
spectrum calculation which cannot be absorbed into redefinitions of the
tree--level couplings. The reading program
should in this case read both the mixing matrix \emph{and} $A_t$ directly
from the spectrum file, without assuming any a priori relation between them.

\subsection{The Model Input File \label{sec:input}}
Here, the user sets up his or her calculation, with low--energy boundary
conditions and SUSY model parameters. If some or all of the
low--energy boundary conditions are not supplied, the spectrum calculator
should use its own defaults for those parameters, passing them on to the
output file, so that the complete set of parameters that has been used for
the calculation is available in the spectrum output. If the spectrum calculator
has hard-coded defaults which the user is not allowed to change,
the parameters that were \emph{actually
used} for the run should be written onto the output file.

The following general structure for the model input file 
is proposed:
\begin{itemize}
\item \ttt{BLOCK MODSEL}:
Program--independent model switches, e.g.\ which model of supersymmetry
  breaking to use.%, whether to include R-parity violation, etc. 
\item \ttt{BLOCK SMINPUTS}: Measured values of SM parameters, used as
  boundary conditions in the spectrum calculation. These are also required
  for subsequent calculations to be consistent with the spectrum
  calculation. 
\item \ttt{BLOCK MINPAR}: 
Input parameters for minimal/default models. No defaults are defined
for this block, and so the user must supply all required parameters. If
\ttt{MINPAR} is not present, then \emph{all} model parameters must be
specified explicitly using \ttt{EXTPAR} below.
\item \ttt{BLOCK EXTPAR}: 
Optional input parameters for non-minimal/non-universal
models. This block may be entirely absent from the input file, in which case
a minimal type of the selected SUSY breaking model will be used.
\end{itemize}
See also the example model input file included in appendix \ref{app:dec}.
\clearpage
\subsection*{\ttt{BLOCK MODSEL}\label{sec:modsel}}
Switches and options for model selection. The entries in
this block should consist of an index, identifying the particular switch in the
listing below, followed by another integer or real number, specifying the
option or value chosen. Switches so far defined are: \\[2mm]
\numentry{1}{Choice of SUSY breaking model. By default, a
minimal type of model will always be assumed. Possible
values are:\\
\snumentry{0}{General MSSM Simulation.} 
\snumentry{1}{(m)SUGRA model.} 
\snumentry{2}{(m)GMSB model.}
\snumentry{3}{(m)AMSB model.}
\snumentry{4}{...}}\\
%\numentry{2}{(Default=1) Choice of input parameter set used to extract the
%  low--scale gauge and (third generation) Yukawa couplings. So far, only one
%  possibility is defined:\\
%\snumentry{1}{$\alpha_{\mrm{em}}^{-1}(m_\Z)^{\mathrm{OS}}, G_F,
%  \alpha_s(m_\Z)^\MSbar, m_\Z, m_b(m_b)^\MSbar, m_t, m_\tau$. }}\\
\numentry{3}{(Default=0) Choice of particle content. This switch is only
  meant as an example and is not yet implemented in any actual code. 
Switches defined could be:\\
\snumentry{0}{MSSM.}
\snumentry{1}{NMSSM.}
\snumentry{2}{...}}\\
\numentry{11}{(Default=1) Number of points for a logarithmically spaced
  grid in $Q$ for which the user wants the spectrum calculator to give output
  for the running parameters. \\ 
\snumentry{1}{one copy of each block containing running parameters will be
  output, at the scale specified by $Q_{\mathrm{max}}$ below.} 
\snumentry{n > 1}{$n$ copies of each block containing running parameters will
  be  output. The smallest scale for the grid 
is normally $m_\Z$ while the maximum scale is set by $Q_{\mathrm{max}}$
below.}}\\   
\numentry{12}{(Default=$M_{\mrm{EWSB}}$) $Q_{\mathrm{max}}$. The largest $Q$
scale at which to give the running parameters if a grid of output scales for
each running parameter block has been requested using the switch above. The
default 
is to give only one copy of each running parameter block, at the scale taken
by the spectrum calculator to perform electroweak symmetry breaking.  This is
often taken to be $Q=M_{\mrm{EWSB}}\equiv\sqrt{m_{\ti{t}_1}m_{\ti{t}_2}}$.}
\numentry{21}{
PDG code for a particle. The running SUSY-breaking mass parameters 
        will be printed out at the pole mass of that particle, in addition 
        to their values at scales given by the grid specified above. 
        Several different entries can be given. } 

\subsection*{\ttt{BLOCK SMINPUTS}\label{sec:sminputs}}
Measured SM parameters, used as boundary conditions for the spectrum
calculation and passed to subsequent calculations for consistency. Note that
some programs have hard-coded defaults for various of these parameters,
hence only a subset may sometimes be available as free inputs.  The 
parameters, as defined in section \ref{sec:smconv}, are:\\[2mm] 
\numentry{1}{$\alpha_\mrm{em}^{-1}(m_{\Z})^{\MSbar}$. Inverse 
 electromagnetic coupling at the $\Z$ pole in the $\MSbar$ scheme  (with 5
 active flavours).} 
\numentry{2}{$G_F$. Fermi constant (in units of $\GeV^{-2}$).}
\numentry{3}{$\alpha_s(m_{\Z})^{\MSbar}$. Strong coupling at
  the $\Z$ pole in the $\MSbar$ scheme (with 5 active flavours).}
\numentry{4}{$m_\Z$, pole mass.}
\numentry{5}{$m_b(m_b)^{\MSbar}$. $b$ quark running mass in the $\MSbar$
  scheme.}
\numentry{6}{$m_t$, pole mass.}
\numentry{7}{$m_\tau$, pole mass.}\\
Please note that $m_b^{\mathrm{pole}}\ne m_b(m_b)^{\MSbar}\ne
m_b(m_b)^{\DRbar}$, see discussions 
in section \ref{sec:smconv} and \cite{Melnikov:2000qh,Baer:2002ek}.

\subsection*{\ttt{BLOCK MINPAR}}
Input parameters for minimal/default SUSY models. 
The interpretation given to the
contents of this block naturally depends on which type of SUSY breaking model
has been selected in block \ttt{MODSEL}. Below are listed how \ttt{MINPAR}
should be filled for mSUGRA, mGMSB, and mAMSB models, and for a general MSSM
setup. All parameters are understood to be \DRbar\ parameters given at the
input scale, \mgut, which by default is the unification scale inferred from
coupling unification. Alternatively, \mgut\ can be given explicitly in block
\ttt{EXTPAR} below. The only exception is $\tan\beta$ which
we define (cf.~section \ref{sec:susypar}) as a \DRbar\ parameter at the scale
$m_\Z$.   

If a non-minimal type of model is desired, these minimal parameter sets may
still be used to form the basis for the spectrum calculation, see
\ttt{EXTPAR} below for details on this. \\[3mm]
\arrdes{mSUGRA models.}
\numentry{1}{$m_0$. Common scalar mass.} 
\numentry{2}{$m_{1/2}$. Common gaugino mass.}
\numentry{3}{$\tan\beta$. Ratio of Higgs vacuum expectation, see section \ref{sec:susypar}.} 
\numentry{4}{sign$(\mu)$. Sign of the bilinear Higgs term in the
  superpotential.}  
\numentry{5}{$A$. Common trilinear coupling.}
\arrdes{mGMSB models.}
\numentry{1}{$\Lambda$. Scale of soft SUSY breaking felt by the low--energy
  sector.} 
\numentry{2}{$\mmess$. Overall messenger scale.}
\numentry{3}{$\tan\beta$. Ratio of Higgs vacuum expectation values, see section \ref{sec:susypar}.}
\numentry{4}{sign$(\mu)$. Sign of the bilinear Higgs term in the
  superpotential.}  
\numentry{5}{$N_{5}$. Messenger index.} 
\numentry{6}{$c_{\mathrm{grav}}$. Gravitino mass factor.}
\arrdes{mAMSB models.}
\numentry{1}{$m_0$. Common scalar mass term.}
\numentry{2}{$m_{3/2}$. Gravitino mass.} 
\numentry{3}{$\tan{\beta}$. Ratio of Higgs vacuum expectation values, see section \ref{sec:susypar}.}
\numentry{4}{sign$(\mu)$. Sign of the bilinear Higgs term in the
  superpotential.}  
\arrdes{Other models.}
\numentry{3}{$\tan{\beta}$. Ratio of Higgs vacuum expectation values, see
  section \ref{sec:susypar}.} 
%\numentry{4}{sign$(\mu)$}
No model--specific standards for inputs for models beyond mSUGRA, mGMSB, and
mAMSB have yet been defined, apart from the non-universality options
available for these models in \ttt{EXTPAR} below. However, as 
long as a code for an alternative SUSY--breaking model adheres to
the \emph{output} standards described in the next section, there should be no
problems in using it with this interface, as long as $\tan\beta$ is still 
provided. 
\clearpage
\subsection*{\ttt{BLOCK EXTPAR}}
Optional input parameters for non-minimal/non-universal
models. This block may be entirely absent from the input file, in which case
a minimal type of the selected SUSY breaking model will be used. When
block \ttt{EXTPAR} is present, the starting
point is still a minimal model with parameters as given in \ttt{MINPAR} but 
with each value present in \ttt{EXTPAR} replacing the 
minimal model value of that parameter, as applicable. If
\ttt{MINPAR} is not present, then \emph{all} model parameters must be
specified explicitly using \ttt{EXTPAR}. All scale-dependent parameters are
understood to be given in the \DRbar\ scheme.\\[3mm]
\numentry{0}{$\mgut$. Input scale for SUGRA, AMSB, and general MSSM models. 
If absent, the GUT scale derived from gauge unification will be used as input
scale. Note that this parameter has no effect in GMSB scenarios
  where the input scale by definition is identical to the messenger scale,
  $\mmess$. A special case is when 
  $Q=M_{\mrm{EWSB}}\equiv\sqrt{m_{\ti{t}_1}m_{\ti{t}_2}}$ is desired as input
  scale, since this scale is not known beforehand. This choice can be invoked by
  giving the special value $\mgut=-1$.
}
\arrdes{Gaugino Masses}
\numentry{1}{$M_1(\mgut)$. $U(1)_Y$ gaugino (Bino) mass.} 
\numentry{2}{$M_2(\mgut)$. $SU(2)_L$ gaugino (Wino) mass.}
\numentry{3}{$M_3(\mgut)$. $SU(3)_C$ gaugino (gluino) mass.} 
\arrdes{Trilinear Couplings}
\numentry{11}{$A_t(\mgut)$. Top trilinear coupling. }
\numentry{12}{$A_b(\mgut)$. Bottom trilinear coupling.}
\numentry{13}{$A_\tau(\mgut)$. Tau trilinear coupling.}
\arrdes{Higgs Parameters\\[1mm]
{--- Only one of the parameter sets ($m_{H_1}^2$,
  $m_{H_2}^2$), ($\mu,m_A^2$), ($\mu,m_{\A^0}$), or ($\mu,m_{\H^+}$) 
should be given, they merely represent different ways of specifying
the same parameters.}} 
\numentry{21}{$m_{H_1}^2(\mgut)$. Down type Higgs mass squared.} 
\numentry{22}{$m_{H_2}^2(\mgut)$. Up type Higgs mass squared. }
\numentry{23}{$\mu(\mgut)$. $\mu$ parameter.}
\numentry{24}{$m_A^2(\mgut)$. Tree--level pseudoscalar Higgs mass
  parameter squared, as defined by eq.~(\ref{eq:mA}).}
\numentry{25}{$\tan\beta(\mgut)$. If present, this value of $\tan\beta$
  overrides the one in \ttt{MINPAR}, and the input scale is taken as $\mgut$
  rather than $m_\Z$.}
\numentry{26}{$m_{\A^0}$. 
  Pseudoscalar Higgs pole mass. May be given
  instead of $m_A^2(\mgut)$.}
\numentry{27}{$m_{\H^+}$. 
  Charged Higgs pole mass. May be given
  instead of $m_A^2(\mgut)$.}
\arrdes{Sfermion Masses}
\numentry{31}{$m_{\ti{e}_L}(\mgut)$. Left 1\st gen.\ scalar lepton mass. }
\numentry{32}{$m_{\ti{\mu}_L}(\mgut)$. Left 2\nd gen.\ scalar lepton mass.}
\numentry{33}{$m_{\ti{\tau}_L}(\mgut)$. Left 3\rd gen.\ scalar lepton mass. }
\numentry{34}{$m_{\ti{e}_R}(\mgut)$. Right scalar electron mass.}
\numentry{35}{$m_{\ti{\mu}_R}(\mgut)$. Right scalar muon mass.}
\numentry{36}{$m_{\ti{\tau}_R}(\mgut)$. Right scalar tau mass.}
\numentry{41}{$m_{\ti{q}_{1L}}(\mgut)$. Left 1\st gen.\ scalar quark mass. }
\numentry{42}{$m_{\ti{q}_{2L}}(\mgut)$. Left 2\nd gen.\ scalar quark mass. }
\numentry{43}{$m_{\ti{q}_{3L}}(\mgut)$. Left 3\rd gen.\ scalar quark mass. }
\numentry{44}{$m_{\ti{u}_{R}}(\mgut)$. Right scalar up mass. }
\numentry{45}{$m_{\ti{c}_{R}}(\mgut)$. Right scalar charm mass. }
\numentry{46}{$m_{\ti{t}_{R}}(\mgut)$. Right scalar top mass. }
\numentry{47}{$m_{\ti{d}_{R}}(\mgut)$. Right scalar down mass. }
\numentry{48}{$m_{\ti{s}_{R}}(\mgut)$. Right scalar strange mass. }
\numentry{49}{$m_{\ti{b}_{R}}(\mgut)$. Right scalar bottom mass. }
%\numentry{20}{$\phi_{\mu}(\mgut)$. Phase of $\mu$ term. }
%\numentry{21}{$\phi_{M_1}(\mgut)$. Phase of $M_1$. }
%\numentry{22}{$\phi_{M_2}(\mgut)$. Phase of $M_2$. }
%\numentry{23}{$\phi_{M_3}(\mgut)$. Phase of $M_2$. }
%\numentry{24}{$\phi_{A_\tau}(\mgut)$. Phase of $A_\tau$. }
%\numentry{25}{$\phi_{A_b}(\mgut)$. Phase of $A_b$.}
%\numentry{26}{$\phi_{A_t}(\mgut)$. Phase of $A_t$.}
\arrdes{Other Extensions}
\numentry{51}{$N_1$ (GMSB only). 
$U(1)_Y$ messenger index (defined as in ref.~\cite{Martin:1996zb}).}
\numentry{52}{$N_2$ (GMSB only). 
$SU(2)_L$ messenger index (defined as in ref.~\cite{Martin:1996zb}).}
\numentry{53}{$N_3$ (GMSB only). 
$SU(3)_C$ messenger index (defined as in ref.~\cite{Martin:1996zb}).}

\subsection{The Spectrum File \label{sec:spectrum}}
%As described above, the MSSM mass and coupling spectrum is specified 
%two types of parameters: those which
%have no explicit scale dependence (the pole masses and the mixing matrices), 
%and the $\DRbar$ running parameters (gauge, trilinear, and Yukawa couplings). 
For the MSSM mass and coupling spectrum, 
the following block names are defined, to be specified further below:
\begin{itemize}
  \item \ttt{BLOCK MASS}: Mass spectrum (pole masses). 
  \item \ttt{BLOCK NMIX}: Neutralino mixing matrix.
  \item \ttt{BLOCK UMIX}: Chargino $U$ mixing matrix.
  \item \ttt{BLOCK VMIX}: Chargino $V$ mixing matrix.
  \item \ttt{BLOCK STOPMIX}: Stop mixing matrix. 
  \item \ttt{BLOCK SBOTMIX}: Sbottom mixing matrix.
  \item \ttt{BLOCK STAUMIX}: Stau mixing matrix.
  \item \ttt{BLOCK ALPHA}: Higgs mixing angle $\alpha$.
  \item \ttt{BLOCK HMIX Q= ...}: $\mu$, $\tan\beta$, $v$, and $m_A^2$ 
at scale $Q$.
  \item \ttt{BLOCK GAUGE Q= ...}: Gauge couplings at scale
    $Q$.  
  \item \ttt{BLOCK MSOFT Q= ...}: Soft SUSY breaking mass parameters at
    scale $Q$. 
  \item \ttt{BLOCK AU, AD, AE Q= ...}: Trilinear couplings at scale $Q$.
  \item \ttt{BLOCK YU, YD, YE Q= ...}: Yukawa couplings at scale $Q$. 
%  \item (\ttt{BLOCK SOFTTERMS Q= ...}: Optional listing of the soft
%  breaking parameters (besides the $A_i$) at the scale $Q$. Warning: this
%  block is  \emph{only} intended for expert users interfacing the spectrum calculator
%  spectra to higher--order calculations. It should \emph{not} be used
%  for tree--level calculations.) 
  \item \ttt{BLOCK SPINFO}: Information from the spectrum calculator. 
\end{itemize}
Note that there should always be at least one empty character between
the \ttt{BLOCK} statement and the block name. For running parameters, 
an arbitrary number of each block may be written, to provide 
parameters at a grid of scales $Q_i$ (in the 
$\overline{\mathrm{DR}}$ scheme). For these blocks,  
the \ttt{Q=} statement should have at least one empty character on
  both sides. See also the example spectrum file included in appendix
  \ref{app:specfile}. 
\subsection*{\ttt{BLOCK MASS}\label{sec:mass}}
Mass spectrum for sparticles and Higgs bosons, signed pole masses.
The standard for each line in the block 
should correspond to the FORTRAN format\begin{center}
\ttt{(1x,I9,3x,1P,E16.8,0P,3x,'\#',1x,A)},\end{center} where the first
9--digit integer should be the PDG code of a particle and the double
precision number its mass. 
\subsection*{\ttt{BLOCK NMIX, UMIX, VMIX, STOPMIX, SBOTMIX,
    STAUMIX}\label{sec:nmix}} 
Mixing matrices, real parts only (CP violation is not addressed by this
Accord at the present stage). 
The standard should correspond to the FORTRAN format\begin{center}
\ttt{(1x,I2,1x,I2,3x,1P,E16.8,0P,3x,'\#',1x,A)}. \end{center}
For a generic mixing matrix $X$,   
the first two integers in the format represent $i$ and $j$ in $X_{ij}$
respectively, and the double precision number $X_{ij}$ itself. Note that
different spectrum calculators may produce different 
overall signs for rows in these matrices, since an overall sign of an
eigenvector does not change physics (see section \ref{sec:mixing} above).
%\subsection*{\ttt{BLOCK NMIXIM, UMIXIM, VMIXIM}\label{sec:nmixim}} 
%Mixing matrices, imaginary parts. See section \ref{sec:nmix} above. If these
%blocks are absent, all imaginary parts will be assumed zero.

\subsection*{\ttt{BLOCK ALPHA}\label{sec:alpha}}
This block only contains one entry, the Higgs mixing angle
$\alpha$ (see definition in Section \ref{sec:mixing}), 
written in the format
\begin{center}
\ttt{(9x,1P,E16.8,0P,3x,'\#',1x,A)}.
\end{center} 

\subsection*{\ttt{BLOCK HMIX Q= ...}\label{sec:hmix}}
$\DRbar$ Higgs parameters at the scale $Q$, cf.~sections \ref{sec:susypar}
and \ref{sec:susybreak}. 
The entries in this block should consist of an 
index, identifying the particular parameter in the listing below, followed by
a double precision number, giving the parameter value. The corresponding
FORTRAN format would be
\begin{center}
\ttt{(1x,I5,3x,1P,E16.8,0P,3x,'\#',1x,A)}.
\end{center} 
So far, the following entries are defined: \\[2mm] 
\numentry{1}{$\mu(Q)$. }
\numentry{2}{$\tan\beta(Q)$. }
\numentry{3}{$v(Q)$. }
\numentry{4}{$m_A^2(Q)$.}

\subsection*{\ttt{BLOCK GAUGE Q= ...}\label{sec:gauge}}
$\DRbar$ gauge couplings at the scale $Q$, cf.~section
\ref{sec:susypar}. 
The entries in this block
should consist of an index, identifying the parameter in the
listing below, the format being equivalent to that of block \ttt{HMIX}
above. The parameters are: \\[2mm]
\numentry{1}{$g'(Q)$.}
\numentry{2}{$g(Q)$.}
\numentry{3}{$g_3(Q)$.}

\subsection*{\ttt{BLOCK MSOFT Q= ...}\label{sec:msoft}}
$\DRbar$ soft SUSY breaking mass parameters at the scale $Q$,
cf.~eqs.~(\ref{eq:v2}) and (\ref{eq:LG}).  The entries in this block
should consist of an index, identifying the parameter in the
listing below, the format being equivalent to that of block \ttt{HMIX}
above and the numbering parallelling that of block \ttt{EXTPAR}. 
The parameters defined are: \\[2mm]
\numentry{1}{$M_1(Q)$.}
\numentry{2}{$M_2(Q)$.}
\numentry{3}{$M_3(Q)$.}
\numentry{21}{$m_{H_1}^2(Q)$.} 
\numentry{22}{$m_{H_2}^2(Q)$.}
\numentry{31}{$m_{\ti{e}_L}(Q)$.}
\numentry{32}{$m_{\ti{\mu}_L}(Q)$.}
\numentry{33}{$m_{\ti{\tau}_L}(Q)$. }
\numentry{34}{$m_{\ti{e}_R}(Q)$.}
\numentry{35}{$m_{\ti{\mu}_R}(Q)$.}
\numentry{36}{$m_{\ti{\tau}_R}(Q)$.}
\numentry{41}{$m_{\ti{q}_{1L}}(Q)$.}
\numentry{42}{$m_{\ti{q}_{2L}}(Q)$.}
\numentry{43}{$m_{\ti{q}_{3L}}(Q)$.}
\numentry{44}{$m_{\ti{u}_{R}}(Q)$.}
\numentry{45}{$m_{\ti{c}_{R}}(Q)$.}
\numentry{46}{$m_{\ti{t}_{R}}(Q)$.}
\numentry{47}{$m_{\ti{d}_{R}}(Q)$.}
\numentry{48}{$m_{\ti{s}_{R}}(Q)$.}
\numentry{49}{$m_{\ti{b}_{R}}(Q)$.}

\subsection*{\ttt{BLOCK AU, AD, AE Q= ...}\label{sec:aude}}
$\DRbar$ soft breaking trilinear couplings at the scale $Q$,
cf.~eq.~(\ref{eq:trilinear}).  
These blocks are indexed like matrices (formatted like block \ttt{NMIX}
above). At present, only the (3,3) component of each of these blocks should
be given, corresponding to $A_t$, $A_b$, and $A_\tau$, respectively. Other
non-zero components would in general introduce mixing in the first and second
generations, a situtation which cannot be handled by the present Accord. This
possibility is, however, left open for future development and/or private
extensions.

\subsection*{\ttt{BLOCK YU, YD, YE Q= ...}\label{sec:yude}}
$\DRbar$ fermion Yukawa couplings at the scale $Q$,
cf.~eq.~(\ref{eq:superpot}). These blocks 
are indexed like matrices (formatted like block \ttt{NMIX} above).  At
present, only the (3,3) component of each of these blocks should be given,
corresponding to the top quark, bottom quark, and tau lepton Yukawa
couplings, respectively. Comments similar to those for the trilinear
couplings above apply.

\subsection*{\ttt{BLOCK SPINFO}\label{sec:spinfo}}
Information from spectrum calculator. The program name and version
number are obligatory. Optional: warnings and error messages from spectrum
calculation, status return codes, regularization and renormalization
prescription etc. The format should be
\begin{center}
\ttt{(1x,I5,3x,A)}.
\end{center}
Entries so far defined are:\\[2mm] 
\numentry{1}{spectrum calculator name (string).}  
\numentry{2}{spectrum calculator version number (string).}
\numentry{3}{If this entry is present, warning(s) were produced by the
spectrum calculator. The resulting spectrum may still be OK. The entry should
contain a description of the problem (string).}  
\numentry{4}{If this entry
is present, error(s) were produced by the spectrum calculator. The resulting
spectrum should not be used. The entry should contain a description of the
problem (string).}  To illustrate, a certain unlucky choice of input
parameters could result in the following form of block \ttt{SPINFO}:
\vspace*{3mm}

\hspace*{1cm}
\begin{minipage}{10cm}
\begin{verbatim}
BLOCK SPINFO
    1    MyRGE
    2    1.0
    3    Charge and colour breaking global minimum
    3    Desired accuracy not quite achieved
    3    LEP2 Higgs bound violated
    4    No radiative electroweak symmetry breaking
    4    Tachyons encountered
\end{verbatim}
\end{minipage}

\subsection{The Decay File \label{sec:decay}}
The decay table for each particle begins with a statement specifying which
particle is decaying and its total width, in the format: 
\begin{verbatim}
#         PDG           Width         
DECAY   1000021    1.01752300e+00   # gluino decays  
\end{verbatim}
The first integer is a PDG particle number, specifying the identity of the
mother of all subsequent lines until the next \ttt{DECAY} or \ttt{BLOCK}
statement (or end-of-file).  The subsequent real number is that particle's
total width.  The end comment contains a human readable translation of the
PDG code.

Every subsequent line contains a decay channel for this mother in the format:
\begin{verbatim}
#          BR         NDA      ID1       ID2  
     4.18313300E-02    2     1000001        -1   # BR(~g -> ~d_L dbar)
     1.55587600E-02    2     2000001        -1   # BR(~g -> ~d_R dbar)
     3.91391000E-02    2     1000002        -2   # BR(~g -> ~u_L ubar)
     1.74358200E-02    2     2000002        -2   # BR(~g -> ~u_R ubar)
...
\end{verbatim}
where the first real number is the fraction of the total width (branching
fraction) into that particular mode, the first integer is the number of
daughters, $N$, and the $N$ following integers are the PDG codes of the $N$
daughters.  The specific FORTRAN formats for the \ttt{DECAY} statement and
the entries in the decay table are, respectively:
\begin{center}
\ttt{('DECAY',1x,I9,3x,1P,E16.8,0P,3x,'\#',1x,A)},
\ttt{(3x,1P,E16.8,0P,3x,I2,3x,$N$(I9,1x),2x,'\#',1x,A)}.\end{center}

A potential pitfall in using these decay tables is how on--shell resonances
inside the physical phase space are dealt with. A prime example, which we will
use for illustration below, is the top decay, $\t\to W^+\b \to \q\qbar'
\b$. There are two dangers here that must be consistently dealt with: 
\begin{enumerate}
\item If both the partial width for $\t\to W^+\b$ and for $\t\to \q\qbar' \b$
  are written on the file, then the on--shell $W$ part of $\t\to \q\qbar'
  \b$ will be counted twice. One solution here is to include only the truly
  off--shell 
  parts in the calculation of partial widths for processes which can occur
  via seuqences of (lower--order) on--shell splittings. In the example above,
  the ($1\to3$)
  partial width $\t\to \q\qbar' \b$ written on the file should thus
  \emph{not} contain the ($1\to2 \otimes 1\to2$) on--shell $W$ contribution.
\item If the on--shell/off--shell rule just described is \emph{not} adhered
  to, another possible source of error becomes apparent. If the full partial
  width $\t\to \q\qbar' \b$ (now including the resonant piece) is written
  in the decay table, the reading program has no immediate way of knowing
  that there is a resonant $W$ inside. By default, it will most likely use a
  flat phase space, in the lack of more differential information; hence there
  would in this example be no $W$ peak in the $\q\qbar'$ invariant mass
  spectrum. On the other hand, if the rule above \emph{is} adhered to, then 
  the resonant $W$ is not part of the $1\to 3$ partial width, and a flat
  phase space is then a reasonable first approximation.
\end{enumerate}

NB: for Majorana particles, modes and charge conjugate modes should
both be written on the file, so that the numbers in the first column
sum up to 1 for any particle.
 
At present, this file is thus only capable of transferring integrated
information, i.e.\ partial widths. For a more accurate population of phase
space (even when all intermediate states are off--shell),
access to differential information is necessary. It should also here be
mentioned that several
programs include options for letting the partial widths be a function
of $\hat{s}$, to account for the resonant shape of the mother. 
We anticipate that these and other refinements can be included with full
backwards 
compatibility, either by continuing to add information on the same line
before the hash mark, or by adding a number of lines beginning with the
'\ttt{+}' character for each decay mode in question, where additional
information concerning that mode can be given.

The file should also contain a block \ttt{DCINFO}, giving information about
the decay calculation program and (optionally) exit status. The format and
entries of this block are identical to that of block \ttt{SPINFO} above.

Note that, if the particle is not known to the reading program, properties
such as electric charge, spin, and colour charge could be added on the
\ttt{DECAY} line. However, for fundamental particles, there is only a limited
set of particles for which all programs concerned should know the properties,
and for hadrons, the spin is encoded in the last digit of the PDG code while
the charge may be calculated from the flavour content which is specified by
the 2 or 3 digits preceding the spin digit of the code.  Therefore, we do not
deem it necessary to adopt a standard for specifying such information at the
present time.

\section{Conclusion}
The present Accord specifies a unique set of conventions together with 
ASCII file formats for model input and spectrum output for most commonly
investigated supersymmetric models, as well as a decay table file format for
use with decay packages. 

With respect to the model parameter input file, mSUGRA, mGMSB, and mAMSB
scenarios can be handled, with some options for non-universality. However,
this should not discourage users desiring to investigate alternative models;
the definitions for the spectrum output file are at present capable of
handling any CP and R--parity conserving supersymmetric model, with the
particle spectrum of the MSSM. Specifically, this includes the so-called SPS
points \cite{Allanach:2002nj}.
 
Also, these definitions are not intended to be static solutions. Great
efforts have gone into ensuring that the Accord may accomodate essentially
any new model or new twist on an old one with minor modifications required
and full backwards compatibility. Planned issues for future extensions of the
Accord are, for instance, to include options for R--parity violation and CP
violation, and possibly to include definitions for an NMSSM. Topics which are
at present only implemented in a few codes, if at all, will be taken up as the
need arises. Handling RPV and CPV should require very minor modifications to
the existing structure, while the NMSSM, for which there is at present not
even general agreement on a unique definition, will require some additional
work. 

Lastly, while SUSY is perhaps the most studied hypothesis of New Physics, it
is by no means the only possible worth investigating. One may well anticipate
that similar sources of confusion and misunderstandings as partly motivated
this Accord can arise for other New Physics models in the future. In this
context, we note that the decay tables defined here are sufficiently general
to require little or no modification to encompass other New Physics models,
while the rest of the Accord provides a general structure that 
may be used as a convenient template for future generalisations.
\\[0.2cm] 
\begin{center}
{\bf Acknowledgements}
\end{center}
The authors are grateful to the organizers of the Physics at TeV
Colliders workshop (Les Houches, 2003) and to the organizers of the
Workshop on Monte Carlo tools for the LHC (MC4LHC, CERN, 2003). The
discussions and agreements reached at those two workshops
constitute the backbone of this writeup.

This work has been supported in part by CERN, by the 
``Collider Physics'' European Network under contract 
HPRN-CT-2000-00149, by the Swiss Bundesamt f\"ur Bildung und
Wissenschaft, and by the National Science Foundation under Grant No.\
PHY99-07949. Research at ANL is supported by the U.S.\ Department of Energy
HEP Division under the contract W-31-109-ENG-38. W.P.~is supported by the
Erwin Schr\"odinger  
fellowship No.~J2272 of the `Fonds zur F\"orderung der wissenschaftlichen
Forschung' of Austria and partly by the Swiss 'Nationalfonds'.
\clearpage
\appendix 
\section{The PDG Particle Numbering Scheme \label{app:pdg}}
Listed in the tables below are the PDG codes for the MSSM particle
spectrum. Codes for other particles may be found in \cite[chp.~33]{Hagiwara:2002fs}. \\[1cm]
\begin{table}[ph]
\captive{SM fundamental particle codes (+ extended Higgs sector).
\protect\label{tab:smcodes} } \\
\vspace{-2ex}
\begin{center}
\begin{tabular}{|c|c||c|c||c|c||c|c|@{\protect\rule{0mm}{\tablinsep}}}
\hline
Code & Name & Code & Name &Code & Name & Code & Name \\
\hline
    1 & $\d$  &  11 & $\e^-$    &    21 & $\g$   & &      \\
    2 & $\u$  &  12 & $\nu_{\e}$&    22 & $\gamma$  & 35 & $\H^0$ \\
    3 & $\s$  &  13 & $\mu^-$   &    23 & $\Z^0$    & 36 & $\A^0$ \\  
    4 & $\c$  &  14 & $\nu_{\mu}$&   24 & $\W^+$    & 37 & $\H^+$ \\
    5 & $\b$  &  15 & $\tau^-$   &   25 & $\hrm^0$     & &  \\
    6 & $\t$  &  16 & $\nu_{\tau}$&      &              &39 & $\mathrm{G}$ (graviton) \\ 
\hline
\end{tabular}
\end{center}
\end{table}
\vspace*{1cm}
\begin{table}[ph!]
\captive{MSSM sparticle codes \protect\label{tab:susycodes} }  \\
\vspace{-2ex}
\begin{center}
\begin{tabular}{|c|c||c|c||c|c||c|c|@{\protect\rule{0mm}{\tablinsep}}}
\hline
Code & Name & Code & Name & Code & Name & Code & Name \\
\hline
1000001 & $\sqd_L$ & 1000011 & $\se_L$ & 1000021 & $\glu$ & & \\
1000002 & $\squ_L$ & 1000012 & $\snu_{{\e}L}$ & 1000022 & $\neut_1$ & 1000035 & $\neut_4$ \\  
1000003 & $\sqs_L$ & 1000013 & $\smu_L$ & 1000023 & $\neut_2$ & &\\
1000004 & $\sqc_L$ & 1000014 & $\snu_{{\mu}L}$ &  1000024 & $\charg_1$ &
1000037 & $\charg_2$ \\ 
1000005 & $\sqb_1$ & 1000015 & $\stau_1$ & 1000025
& $\neut_3$ & & \\
1000006 & $\sqt_1$ & 1000016 & $\snu_{{\tau}L}$ & & &1000039 & $\grav$ (gravitino) \\
\hline 
2000001 & $\sqd_R$ & 2000011 & $\se_R$ & & & & \\
2000002 & $\squ_R$ & 2000012 & $\snu_{{\e}R}$ & & & &\\
2000003 & $\sqs_R$ & 2000013 & $\smu_R$ & & & &\\
2000004 & $\sqc_R$ & 2000014 & $\snu_{{\mu}R}$ & & & &  \\
2000005 & $\sqb_2$ & 2000015 & $\stau_2$ & & & & \\
2000006 & $\sqt_2$ & 2000016 & $\snu_{{\tau}R}$ & & & & \\
%  \\
\hline
\end{tabular}
\end{center}
\end{table}
\clearpage
%%\section{Convention Comparisons \label{app:conventions}}
%%\begin{table}[h!]
%%\begin{center}
%%\begin{tabular}{ccp{10cm}}
%%LHA3 & G\&H & Parameter description\\[1mm]\toprule
%%$Y$ & $\frac12 y$ & Hypercharge \\ \cmidrule{1-3}
%%$\epsilon_{ab}$ & $\varepsilon_{ab}$ & Levi-Civita tensor \\ \cmidrule{1-3}
%%$L$ & $\hat{L}$ & Left leptonic supermultiplet \\
%%$\bar{E}$ & $\hat{R}$ & Right (c.c.) leptonic supermultiplet \\
%%$Q$ & $\hat{Q}$ & Left quark supermultiplet \\
%%$\bar{U},\bar{D}$ & $\hat{U},\hat{D}$ & Right (c.c.) up and down quark supermultiplets \\
%%$H_{1,2}$ & $\hat{H}_{1,2}$ & $Y=-1/2$ and $Y=1/2$ higgs supermultiplets \\
%%\cmidrule{1-3}
%%$Y_E,Y_D,Y_U$ & $\sqrt{2}(f,f_1,f_2)?$ & Lepton, down quark, and up quark Yukawas\\ 
%%$\mu$ & $-\mu$ & \\
%%$A_U,A_D,A_E$ & $-(A_u,A_d,A_e)$ & Soft SUSY--breaking trilinear couplings\\
%%\cmidrule{1-3}
%%... & ... & ...\\
%%\bottomrule
%%\end{tabular}
%%\caption{The conventions used here as compared to those of
%%  \cite{Gunion:1986yn}.\label{tab:conventions}} 
%%\end{center}
%%\end{table}
\section{1-Loop Translations from \DRbar\ to 
\MSbar\label{app:translations}}
All formulae in this appendix are obtained from
\cite{Martin:1993yx} and are valid for parameters calculated at 1 loop in
either of the two schemes. 

At the scale $\mu$, the \MSbar\ gauge coupling of 
the gauge group $G_i$ (i.e.~the coupling appearing in the interaction
vertices of the corresponding gauge bosons) is related to the \DRbar\ one by:
\begin{equation}
g_{i,\MSbar} = g_{i,\DRbar} \left( 1-\frac{g_i^2}{96\pi^2} C(G_i)\right)~,
\end{equation}
where 
the choice of renormalization scheme for $g_i$ in the 1--loop correction
piece is irrelevant, since it amounts to a 2--loop effect, and
$C(G_i)$ is the quadratic Casimir invariant for the adjoint
representation of the gauge group in question.\footnote{It should be
noted that the couplings of gauginos to scalars are identical to the
gauge couplings by virtue of supersymmetry. This requires the
introduction of additional counter terms in the \MSbar~scheme
\cite{Martin:1993yx} in order to restore this equality. Analogous
additional counter terms arise for the quartic scalar couplings which
are related to the Yukawa and gauge couplings due to supersymmetry.}

For the running Yukawa couplings (between the scalar $\phi_i$ and the
two chiral fermions $\psi_j$ and $\psi_k$), the translation is:
\begin{equation}
Y^{ijk}_{\MSbar} = Y^{ijk}_{\DRbar}\left(1+\sum_{a=1}^3
\frac{g_a^2}{32\pi^2}\left[ C_a(r_j) - 2C_a(r_i) + C_a(r_k) \right]\right)~,
\end{equation}
where $a$ runs over the SM gauge groups. Note that while the \DRbar\
Yukawas are totally symmetric, this is not the case for the \MSbar\
ones. The same relation can also be used to derive the translation of
fermion masses coming from a quadratic term in the superpotential by
taking the scalar field as a dummy field with $C(r_i)=0$ and identifying
$C(r_j)=C(r_k)$. This applies e.g.\ to the Higgs mixing parameter $\mu$.

The soft supersymmetry--breaking parameters differ in the two schemes,
too. The relation between the gaugino masses is given by
\begin{equation}
M_{i,\MSbar} = M_{i,\DRbar} \left( 1+\frac{g_i^2}{16\pi^2} C(G_i)\right)~,
\end{equation}
Finally, none of the other supersymmetry--breaking couplings (as written
in component rather than superfield notation) differ between the two
schemes. In particular, this applies to the soft breaking trilinear
couplings and the scalar masses, provided one uses the
modified \DRbar~scheme, as presented in \cite{Jack:1994rk}.
\clearpage
\section{Tree--level Mass Matrices \label{app:mixing}}
The following gives a list of tree--level mass matrices, as they appear 
in the conventions adopted in this article, see Section
\ref{sec:conventions}. Note that  
these expressions \emph{are not} normally the ones used for actual
calculations in the spectrum calculators, since most codes on the market today
include higher order corrections which are absent below.

The neutralino mass matrix appearing in eq.~(\ref{eq:neutmass}) is:
\begin{equation}
{\cal M}_{\tilde\psi0} \ =\ \left(\begin{array}{cccc} M_1 & 0 &
-m_\Z\cos\beta \sin\theta_W & m_\Z\sin\beta \sin\theta_W \\ 0 & M_2 & m_\Z\cos\beta \cos\theta_W &
-m_\Z\sin\beta \cos\theta_W \\ -m_\Z\cos\beta \sin\theta_W & m_\Z\cos\beta \cos\theta_W & 0 & -\mu \\
m_\Z\sin\beta \sin\theta_W & -m_\Z\sin\beta \cos\theta_W & -\mu & 0
\end{array} \right)~, \label{eq:mchi0}
\end{equation}
and the chargino mass matrix
appearing in eq.~(\ref{eq:chargmass}): 
\begin{equation}
{\cal M}_{\tilde{\psi}^+} =
\left( \begin{array}{cc}
M_2 & \sqrt{2}m_\W\sin\beta \\
\sqrt{2}m_\W\cos\beta & \mu
\end{array}
\right)~.
\end{equation}
For the sfermions, the mixing matrices for $\tilde{t}$, $\tilde{b}$, and
$\tilde{\tau}$ respectively, appear in the L--R basis as:
\begin{equation}
\label{eq:mstop} \left(\begin{array}{cc}
m_{\tilde q_{3L}}^2 + {m_{t}^2} + (\half - \frac{2}{3} 
\sin^2\theta_W)m_\Z^2\cos{2\beta} &
m_{t}\left(A_t-\mu\cot\beta\right)\\
m_{t}\left(A_t-\mu\cot\beta\right) & m^2_{\tilde t_R} +
m_{t}^2 + \frac{2}{3} \sin^2\theta_W m_\Z^2\cos{2\beta}
\end{array}\right)~,
\end{equation}
\begin{equation}
\label{eq:msbot}  \left(\begin{array}{cc}
m_{\tilde q_{3L}}^2 + m_{b}^2 - (\half - \frac{1}{3} 
\sin^2\theta_W)m_\Z^2\cos{2\beta} &
m_{b}\left(A_b-\mu\tan\beta\right)\\
m_{b}\left(A_b-\mu\tan\beta\right) & m^2_{\tilde b_R} +
m_{b}^2 -\frac{1}{3} \sin^2\theta_W m_\Z^2\cos{2\beta}
\end{array}\right)~,
\end{equation}
\begin{equation}
\label{eq:mstau}  \left(\begin{array}{cc}
m_{\tilde \tau_L}^2 + m_{\tau}^2 - (\half - \sin^2\theta_W)m_\Z^2\cos{2\beta} &
m_{\tau}\left(A_\tau-\mu\tan\beta\right)\\
m_{\tau}\left(A_\tau-\mu\tan\beta\right) & m^2_{\tilde \tau_R} +
m_{\tau}^2 - \sin^2\theta_W m_\Z^2\cos{2\beta}
\end{array}\right),
\end{equation}
where we use $m_{\tilde q_{3L}}$ to denote 
the 3\rd~generation left squark mass.
\clearpage
\section{Examples \label{app:examples}}
\subsection{Example Input File \label{app:dec}}
In the example below, the user has not
entered boundary values for the electroweak couplings, nor have 
the $\Z$ and $\tau$ masses been supplied.
On running, the spectrum calculator should thus use its own
defaults for these parameters and pass everything on to the output.\\[4mm]
\hrule\vspace*{0.0cm}
\input{example.par}
\hrule\vspace*{1cm}
\subsection{Example Spectrum File \label{app:specfile}}
The spectrum file produced by the above input file should look something 
like the following:\\[4mm]
\hrule\vspace*{0.0cm}
\input{example.spc}
\hrule\vspace*{1cm}
\subsection{Example Decay File}
For brevity, the model input and spectrum information is omitted here. See
the examples above.  \\
\hrule\vspace*{0.1cm}
\begin{verbatim}
# SUSY Les Houches Accord 1.0 - example decay file
# Info from decay package
Block DCINFO          # Program information
     1    SDECAY       # Decay package
     2    1.0          # version number
#         PDG           Width                  
DECAY   1000021    1.01752300e+00   # gluino decays  
#          BR         NDA      ID1       ID2  
     4.18313300E-02    2     1000001        -1   # BR(~g -> ~d_L dbar)
     1.55587600E-02    2     2000001        -1   # BR(~g -> ~d_R dbar)
     3.91391000E-02    2     1000002        -2   # BR(~g -> ~u_L ubar)
     1.74358200E-02    2     2000002        -2   # BR(~g -> ~u_R ubar)
     4.18313300E-02    2     1000003        -3   # BR(~g -> ~s_L sbar)
     1.55587600E-02    2     2000003        -3   # BR(~g -> ~s_R sbar)
     3.91391000E-02    2     1000004        -4   # BR(~g -> ~c_L cbar)
     1.74358200E-02    2     2000004        -4   # BR(~g -> ~c_R cbar)
     1.13021900E-01    2     1000005        -5   # BR(~g -> ~b_1 bbar)
     6.30339800E-02    2     2000005        -5   # BR(~g -> ~b_2 bbar)
     9.60140900E-02    2     1000006        -6   # BR(~g -> ~t_1 tbar)
     0.00000000E+00    2     2000006        -6   # BR(~g -> ~t_2 tbar)
     4.18313300E-02    2    -1000001         1   # BR(~g -> ~dbar_L d)
     1.55587600E-02    2    -2000001         1   # BR(~g -> ~dbar_R d)
     3.91391000E-02    2    -1000002         2   # BR(~g -> ~ubar_L u)
     1.74358200E-02    2    -2000002         2   # BR(~g -> ~ubar_R u)
     4.18313300E-02    2    -1000003         3   # BR(~g -> ~sbar_L s)
     1.55587600E-02    2    -2000003         3   # BR(~g -> ~sbar_R s)
     3.91391000E-02    2    -1000004         4   # BR(~g -> ~cbar_L c)
     1.74358200E-02    2    -2000004         4   # BR(~g -> ~cbar_R c)
     1.13021900E-01    2    -1000005         5   # BR(~g -> ~bbar_1 b)
     6.30339800E-02    2    -2000005         5   # BR(~g -> ~bbar_2 b)
     9.60140900E-02    2    -1000006         6   # BR(~g -> ~tbar_1 t)
     0.00000000E+00    2    -2000006         6   # BR(~g -> ~tbar_2 t)
\end{verbatim}
\hrule\vspace*{0.6cm}

\bibliography{accord29}
\end{document}